# Comparative Study of Magnetic Moment of Leptons in Hot and Dense Media

Samina Masood* and Holly Mein
Department of Physical and Applied Sciences,
University of Houston-Clear Lake, Houston TX 77058
*E.mail: masood@uhcl.edu

## Abstract

We study the magnetic moment of leptons in extremely hot universe and superdense media of stars at high temperatures. Anomalous magnetic moment of charged leptons is inversely proportional to its mass, whereas, the induced dipole moment of neutral leptons is directly proportional to their mass. Neutral massive point particles exert nonzero magnetic moment as a higher order effect, which is smaller than the anomalous magnetic moment of a charged particle of their same flavor partner. All leptons acquire some extra mass due to their interaction with the medium and affect the magnetic moment accordingly. We compare the contribution to the magnetic moment of various leptons due to their temperature and chemical potential dependent masses. It is shown that the magnetic moment contributions are non-ignorable for lighter leptons and heavy neutrinos. These calculations are very important to study particle propagation in the early universe and in superdense stellar media.

## 1. Introduction

Magnetic moment is expressed in terms of charge and mass of particles. Perturbative changes in the physical mass [1-5] lead to the change in magnetic moment [6-7], which is calculated using the renormalization techniques of quantum electrodynamics (QED) in real-time formalism. This perturbative contribution is identified as the anomalous magnetic moment [8-11]. Contribution of the statistical background [12-16] depends on temperature and chemical potential of the medium, usually expressed in t units of mass of the corresponding lepton. We use the results obtained by using the renormalization scheme of QED in real-time formalism. The magnetic moment of leptons is significantly modified due to the light masses of leptons in statistical background of extremely hot and dense systems. Significance of the change in mass depends on the bare masses of leptons, contributing differently for lighter leptons and heavier ones. It also depends on the mass and density of the statistical conditions of the background.

A comparative study of thermal corrections to the lepton magnetic moment shows that the significance of thermal contributions to the magnetic moment of different leptons in the very early universe depends on the temperature of the universe. For comparison, we focus near the primordial nucleosynthesis temperature. This temperature is taken to be equivalent to electron mass, which is around $10^{10}$ K. Stellar cores have very high temperatures and may have high enough density for which chemical potential may not be ignorable. It has been found earlier that the particles' masses were growing quadratically with temperature in the early universe [1]. The change in mass affects the corresponding value of magnetic moments of leptons [6]. Anomalous magnetic moment is inversely proportional to the mass, whereas the neutrino dipole moment is directly proportional to neutrino mass. For this purpose, the Dirac type neutrino in the minimally extended standard model is considered to assign tiny mass to Dirac type neutrino. It allows to add right-handed neutrinos to the standard model, as a singlet. In this minimally extended standard model, neutrinos satisfy the Dirac equation, just like a charged lepton regardless of their light mass. This Dirac type neutrino exhibits an induced magnetic moment due to its interaction with the relevant charged leptons, which is called the induced magnetic moment. However, the effect of temperature and density on the neutrino appears as a higher order effect, being not as significant as for charged leptons. Statistical corrections to the anomalous magnetic moment of charged leptons come through the radiative corrections to lepton masses or selfmass of leptons. On the other hand, the magnetic moment of neutrinos is induced through the interaction of the neutrino with the charged leptons of the corresponding flavor as well as the W-boson, having a very large mass. as compared to electron. These interactions are shown in Figures (1) and (2).

Renormalization of QED suggest that thermodynamic corrections to electric charge or electromagnetic couplings [6,17,18] are ignorable as compared to similar corrections to mass in the relevant statistical conditions of the early universe, which are ignored in present analysis. We will restrict ourselves to the first order corrections only as they give the leading order contributions in renormalization theories.

In the next section, we discuss the contributions of the lepton magnetic moments in vacuum for different type of leptons. Section 3 is comprised of the comparison of statistical contributions to the magnetic moment of various leptons. These contributions depend on the ratio of statistical parameters with the mass of leptons. These calculations play a crucial role in the detailed understanding of particle dynamics of the early universe and cores of superdense stars. The discussion of the results is included in the last section.

2. **Magnetic Moment of Leptons in Vacuum**

The spin of electron in an atom is added as a perturbative term to the Hamiltonian with the constant magnetic field B as $H_B = \mu_\ell . B$, where $\mu_\ell$ is referred to as the anomalous magnetic moment and is expressed in terms of mass and charge of the particle. The magnetic moment of electron is used as a unit of magnetic moment and is called Bohr magneton $\mu_B$.

$$\mu_e = \frac{e\hbar}{2m_e} \equiv \mu_B \qquad (1)$$

The magnetic moment of a charged lepton flavor ` $\ell$ ' (with $\ell$ = e, μ, τ) is then expressed in units of Bohr magneton, $\mu_B$ for $m_e$, the bare mass of electron and $m_\ell$ the mass of leptons. The magnetic moment is being calculated as $a_\ell$ for flavor ` $\ell$ ' in units of Bohr magneton such that the angular momentum of charged lepton $\ell$ is written as

$$\mu_\ell = \frac{e\hbar}{2m_\ell} = \frac{m_e}{m_\ell}\mu_B \qquad (2)$$

Anomalous magnetic moment is calculated using the renormalization scheme in quantum field theory, which gives an additive contribution to the magnetic moment ($\frac{\alpha}{2\pi}$) in terms of selfmass of leptons, in units of Bohr magneton. This is a general scheme of calculation and is adopted in Ref. [19] as well. It can be easily seen that the magnetic moment is expected to be inversely proportional to the mass of lepton. Heavy particles naturally have lower magnetic moment as compared to the lighter ones with the same charge. Total first order radiative corrections to the magnetic moment of charged lepton with flavor ℓ for leptons, µ_ℓ, can then be expressed, as:

$$\mu_\ell = \frac{\alpha}{2\pi} - \frac{2}{3}\frac{\delta m_\ell}{m_\ell} \qquad (3)$$

The selfmass of lepton $\frac{\delta m_\ell}{m_\ell}$ in Eq.(3) correspond to the first order selfmass corrections as an additive term and gives the radiative corrections to the magnetic moment of charged lepton µ_ℓ and is calculated by using the renormalization scheme of QED. The substitution of the corresponding selfmass value helps to evaluate the correct values of anomalous magnetic moment of a given species of charged leptons at high energy. The anomalous magnetic moment is a local effect of highly energetic system that is associated with the weak magnetic fields of the revolving charges.

**Magnetic Moment of Neutrino:**

Neutrinos are massless neutral particles and do not interact electromagnetically or exhibit intrinsic magnetic moment. However, they can acquire induced magnetic moment due to their interaction with virtual charges and leptons of the same flavor in the medium. The A magnetic moment is induced by the virtually interacting charged lepton and is a distinct feature of neutrinos, which are very light in mass and are electrically neutral. However, neutrino mass is described in different extensions of the standard model differently.

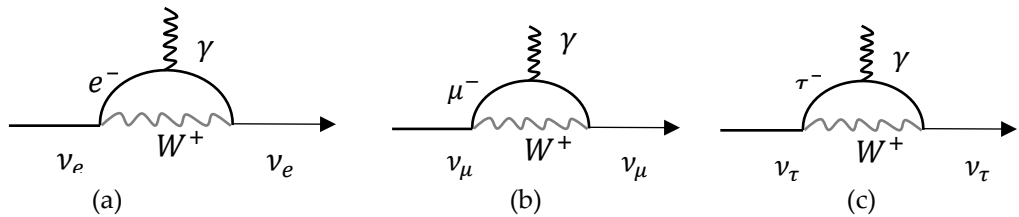

**Fig.1:** Bubble diagrams for different flavors of neutrinos in the minimal standard model with lepton favor conservation. (a) corresponds to electron-neutrino (b) muon+-neutrino and (c) tau-neutrino.

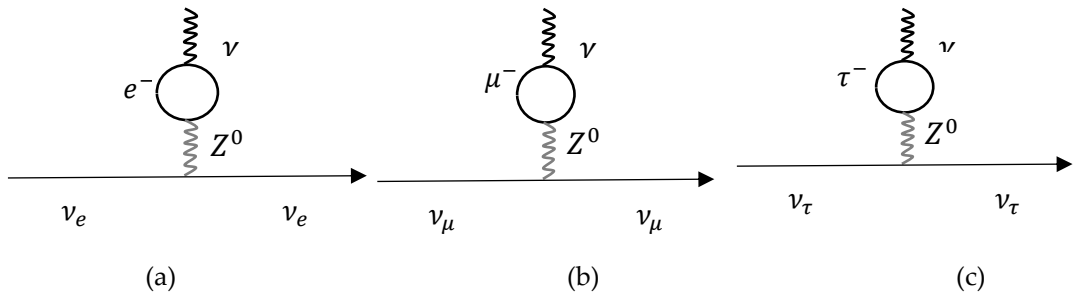

(a)  (b)  (c)

**Fig.2:** Tadpole diagrams for different flavors of neutrinos in the minimal standard model. with lepton favor conservation. (a) corresponds to electron-neutrino (b) muon-neutrino and (c) tau-neutrino

The induced magnetic moment of Dirac neutrino during its virtual interaction with the charged lepton loops up to the one loop level is calculated using bubble diagrams of Figs, (1). The corresponding contributions from tadpole diagrams of Figs, (2) are ignorable due to the neutral current interaction of fermion loops which is suppressed by Z boson mass. Fujikawa and Schrock [8] calculated the magnetic moment of the Dirac neutrino in vacuum, for the first time. These calculations gave the magnetic moment in terms of the mass of neutrino (expressed in electron volt), and are substituted in units of $\mu_B$:

$$\mu_\ell^D = \frac{3eG_F m_{\nu_\ell}}{8\sqrt{2}\pi^2} \approx 3.2 \times 10^{-19} \left(\frac{m_{\nu_\ell}}{1eV}\right) \mu_B \qquad (4)$$

$$G_F = \frac{\sqrt{2}}{8} \frac{g^2}{m_W^2 c^4} = 1.166 \times 10^{-5} GeV^{-2} \qquad (5)$$

$G_F$ is the Fermi Coupling Constant in the weak interaction and $m_{\nu_\ell}$ is the mass of the corresponding flavor of neutrino. The induced magnetic moment is obviously much smaller than the magnetic moment of the corresponding charged lepton of the same flavor. The coupling of W boson with the magnetic field is ignored due to the mass suppression, for the calculation of the leading order contributions.

We generalize Eq. (4) to calculate the magnetic moment of different neutrino species in vacuum. These may be different from the masses that are used in previous papers on the magnetic dipole moment of the neutrino. This is because finding the exact values of the neutrino masses is not easy because their direct measurement is not an easy task and the existing mass limits due to indirect measurements depend on the chosen models. We generalize the Lee and Schrock's results of Ref. [19] for the magnetic moment of an electron neutrino in a vacuum to all neutrino flavors using:

$$\mu_{\nu_\ell}^0 = \frac{3eG_F m_{\nu_\ell}}{8\sqrt{2}\pi^2} = \frac{3m_e G_F m_{\nu_\ell}}{4\sqrt{2}\pi^2} \ (\mu_B) \approx 3.2 \times 10^{-19} \left(\frac{m_{\nu_\ell}}{1eV}\right) \mu_B \qquad (6)$$

Here, $m_{\nu_\ell}$ denotes the mass of neutrino of flavor $\ell$. It should be noticed that the charged lepton magnetic moment is inversely proportional to lepton mass, whereas neutrino magnetic moment is simple proportional to its mass. Magnetic moment of each neutrino flavor can be evaluated from Eq.(6). The statistical contributions to induced magnetic moment in terms of temperature and density will be calculated and compared later in this paper.

The anomalous magnetic moment of leptons in relevant ranges of temperatures and chemical potential are also calculated and compared with the corresponding vacuum values. Magnetic fields in most cases can be considered to be constant and too low to show any measureable effect on the propagation of particles in a medium. We therefore calculate the effects when chemical potential is greater than temperature, which is larger than the mass of the corresponding lepton flavor.

### 3. Magnetic Moments at High Temperatures and Densities

The renormalization scheme of QED has been used in real-time formalism showing the order-by-order cancellation of singularities (KLN-theorem [20-21]). The contribution of the thermal background can be evaluated in real-time formalism [1-7]. The renormalized mass of the leptons can also be described as physical mass of leptons, which can be written as:

$$m_\ell^{phys} = m_\ell + \delta m_\ell \qquad (7)$$

The lepton mass $m_\ell$ represents the bare mass of lepton with the flavor $\ell$. A straightforward generalization of the calculation of electron self-mass (see for example [5]) for various lepton flovors is done in reference to the temperature expressed in units of the same lepton mass $m_\ell$. The first order temperature correction for the electron mass at a finite temperature [5-10] is equal to:

$$\frac{\delta m_\ell}{m_\ell} \simeq \frac{1}{2m_\ell^2}\{(m_\ell^{phys})^2 - m_\ell^2\})$$

$$\simeq \frac{\alpha\pi T^2}{3m_\ell}\left[1 - \frac{6}{\pi^2}c(m_\ell\beta)\right] + \frac{2\alpha}{\pi}\frac{T}{m_\ell}a(m_\ell\beta) - \frac{3\alpha}{\pi}b(m_\ell\beta) \qquad (8)$$

Whereas it can be easily modified to incorporate density effect of the medium by measuring the chemical potential of the relevant species in terms of its rest mass. Selfmass of electron at $T > m_\ell > \mu_\ell$ in real-time formalism can be written in terms of temperature and chemical potential as [5]

$$\frac{\delta m}{m} \approx \frac{\alpha\pi T^2}{3m^2}\left[1 - \frac{6}{\pi^2}c(m\beta,\mu)\right] + \frac{2\alpha}{\pi}\frac{T}{m}a(m\beta,\mu) - \frac{3\alpha}{\pi}b(m\beta,\mu). \qquad (9)$$

With m, the corresponding lepton mass and abc functions are defined as:

$$a(m\beta,\pm\mu) = \ln(1 + e^{-\beta(m\pm\mu)}) \qquad (10a)$$

$$b(m\beta,\pm\mu) = \sum_{n=1}^{\infty}(-1)^n e^{\mp\beta\mu}Ei(-nm\beta) \qquad (10b)$$

$$c(m\beta,\pm\mu) = \sum_{n=1}^{\infty}(-1)^n \frac{e^{-n\beta(m\pm\mu)}}{n^2} \qquad (10c)$$

The electron mass $m_e$ is replaced by $m(1 + \frac{\delta m}{m})$ in the relevant range of temperature and chemical potential in a medium, corresponding to $m_\ell$. Equation (8) gives the selfmass of highly energetic leptons for a known flavor) for relevant ranges of temperatures and densities and can be compared to the mass of leptons.

For $\alpha = \frac{e^2}{4\pi\hbar c} \approx \frac{1}{137}$, the coefficients of $a(m_\ell\beta)$, $b(m_\ell\beta)$ and $c(m_\ell\beta)$ contribute differently at different temperatures (for ignorable chemical potential value), corresponding to the lepton flavors $\ell$. At low temperatures these coefficients are unimportant and can be dropped off, giving selfmass to be:

$$\frac{\delta m_\ell}{m_\ell} \xrightarrow{T \ll m_\ell} \frac{\alpha\pi T^2}{3m_\ell^2} \qquad (11a)$$

The coefficients `a' and `b' are very small and ignorable at high temperature, and `c' is summed to be $-\frac{\pi^2}{12}$, for large T giving:

$$\frac{\delta m_\ell}{m_\ell} \xrightarrow{T \gg m_\ell} \frac{\alpha\pi T^2}{2m_\ell^2} \qquad (11b)$$

Eq. (8) is relevant for calculations pertaining to primordial nucleosynthesis which occurs near the electron mass temperature. Eqs. (8-11) are used to estimate thermal contributions to the lepton mass and magnetic moment of all leptons in the minimally extended standard model. Additionally, the above equations help to determine thermal contributions to all leptons expressed in units of Bohr magneton.

The radiative corrections to the lepton selfmass in a thermal background of fermions and bosons significantly contributes to the magnetic moment of leptons. It has been previously shown [13] that the selfmass of particles contribute to the anomalous magnetic moment $a_a$ (in units of $\mu_B$) of electron as:

$$a_a = \frac{\alpha}{2\pi} - \frac{2}{3}\frac{\delta m_e}{m_e}$$

Which can be generalized in a straightforward manner to anomalous magnetic of moment of any lepton flavor in the units of corresponding flavor.

$$a_\ell^a = \frac{\alpha}{2\pi} - \frac{2}{3}\frac{\delta m_\ell}{m_\ell} \qquad (12)$$

The renormalizability of QED establishes the fact that one-loop thermal corrections dominate over higher order corrections. In this paper, we restrict ourselves to the first order thermal corrections only. Therefore, the one-loop thermal corrections to the magnetic moment of leptons in terms of the corresponding coefficients $a(m_\ell\beta)$, $b(m_\ell\beta)$ and $c(m_\ell\beta)$ [3] are given as:

$$a_\ell^\beta = -\frac{2\alpha\pi T^2}{9m_\ell}\left[1 - \frac{6}{\pi^2}c(m_\ell\beta)\right] + \frac{2\alpha}{\pi}\frac{T}{m_\ell}a(m_\ell\beta) - \frac{3\alpha}{\pi}b(m_\ell\beta) \qquad (13)$$

Whereas the net values of lepton magnetic moment with thermal corrections is:

$$a_\ell^\beta = \left(\frac{\alpha}{2\pi} - \frac{2}{3}\frac{\delta m_\ell}{m_\ell}\right)\left(\frac{m_e}{m_\ell}\right)\mu_B \qquad (14)$$

At temperatures sufficiently below the lepton mass, thermal contributions of the corresponding fermions are stifled in the standard model and only photons contribute to the temperature dependent corrections. However, for T<< $m_\ell$, Eq. (11) leads to:

$$a_\ell^\beta = \frac{\alpha}{2\pi} - \frac{2}{9}\frac{\alpha\pi T^2}{m_\ell^2} \qquad (15a)$$

And, for T >> $m_\ell$:

$$a_\ell^\beta = \frac{\alpha}{2\pi} - \frac{1}{3}\frac{\alpha\pi T^2}{m_\ell^2} \qquad (15b)$$

Thermal contributions to the magnetic moment of electron-neutrino before nucleosynthesis [21-22] at T >> $m_e$ is:

$$a_e = 1.17 \times 10^{-3}\left(1 - \frac{2\pi^2 T^2}{3m_e^2}\right)\mu_B \qquad (16a)$$

Whereas, for T << $m_e$, the magnetic moment value comes out to be:

$$a_e = 1.17 \times 10^{-3}\left(1 - \frac{4\pi^2 T^2}{9m_e^2}\right) \qquad (16b)$$

We see this thermal contribution is greater than the corresponding vacuum value of the magnetic moment at high temperatures causing a flip in the magnetic moment.

Muons and tauons have similar behavior as electrons in thermal background. In this case, the photons interact through muon and tauon loops instead of electron loop due to the individual flavor conservation. For this purpose, we generalize the existing results of electron to all lepton flavors in the early universe. However, the temperature correspondence with the lepton mass will vary with the lepton flavor and is mentioned in Tables 1 and 2.

Thermal contributions to the magnetic moment of neutrinos are calculated from the bubble diagrams of Fig. (1), and they are generalized to all lepton flavors as:

$$a_{\nu_\ell} = \frac{T^2 G_F m_e m_{\nu_l}}{12 M^2} \mu_B \qquad (17)$$

Thermal contributions to the magnetic moment of neutrino at the nucleosynthesis temperatures are given below to show that this contribution is larger for heavier neutrinos as compared to the lighter ones, in the standard model ignoring its interaction with the medium.

$$a_{\nu_e} = \frac{T^2 G_F m_e m_{\nu_e}}{12 M^2} \mu_B \approx 2.92 \times 10^{-16} \mu_B \qquad (18a)$$

$$a_{\nu_\mu} = \frac{T^2 G_F m_e m_{\nu_\mu}}{12 M^2} \mu_B \approx 1.91 \times 10^{-11} \mu_B \qquad (18b)$$

$$a_{\nu_\tau} = \frac{T^2 G_F m_e m_{\nu_\tau}}{12 M^2} \mu_B \approx 1.55 \times 10^{-9} \mu_B \qquad (18c)$$

It is also worth-mentioning that we include thermal corrections to mass and the magnetic moment, but keep the charge and the coupling constant independent of temperature. It will not significantly affect the results as the thermal corrections to charge and the QED coupling constant are much smaller than the corresponding corrections to mass and the magnetic moment even at nonignorable densities.

4. Discussion

Leptons were among the point particles that are produced in the beginning of the universe. A detailed understanding of the behavior of particles in the early universe is expected to resolve unsolved mysteries of the universe. High temperatures of that early universe must have played a crucial role in particle behavior. Lepton mass and charge are known to be modified in QED plasma during this time, through vacuum polarization [24]. Electromagnetic properties of particles in the early universe are also determined from mass, charge and the magnetic moment of leptons.

Implications of thermal corrections to the magnetic moment are relevant in the early universe especially at very high temperatures. The magnetic moment of charged leptons is inversely proportional to the square of its mass, so the heavier leptons have much smaller magnetic moment compared to electrons (Eqs. (11)). Thermal contributions to magnetic moment of neutrinos are proportional to neutrino masses (Eq.17) and thermal corrections to heavier neutrinos contribute more significantly than lighter ones. However, the heavier neutrinos can only exist at higher energies.

Thermal contribution to lepton magnetic moment is subtracted from the magnetic moment in vacuum reducing its net value with the increase in temperature. It is demonstrated by the fact that thermal contribution to electron magnetic moment is around 0.7% whereas the contributions to the muon or tauon is much smaller and ignorable, as shown in Table 1.

Table 1: Magnetic Moment of Charged Leptons around Nucleosynthesis in the Universe

| Charged Leptons | Mass (eV) | Corresponding Temperature (K) | Magnetic Moment at T=0 ($\mu_B$) | Thermal Contribution at T= $m_e$ |
|---|---|---|---|---|
| $e$ | $0.511 \times 10^6$ | $0.592 \times 10^{10}$ | 1 | $-7.6 \times 10^{-3} \mu_B$ |
| $\mu$ | $105.65 \times 10^6$ | $0.122 \times 10^{13}$ | $4.8 \times 10^{-3}$ | $-1.19 \times 10^{-7} \mu_B$ |
| $\tau$ | $1776.82 \times 10^6$ | $0.206 \times 10^{14}$ | $2.8 \times 10^{-4}$ | $-4.2 \times 10^{-11} \mu_B$ |

At higher temperatures near the tauon mass, the magnetic moment of the higher mass particles will not remain ignorable. We have plotted the magnetic moment of electron, muon and tau leptons as a function of temperature in Figs. (3). Temperature dependence is similar whereas the magnitude of thermal contribution is suppressed with the square of mass. All of these magnetic moments are plotted in units of Bohr magneton.

**Fig. 3a: Magnitude of Electron Magnetic Moment (in units of $\mu_B$) vs $\left(\frac{T}{m}\right)^2$**

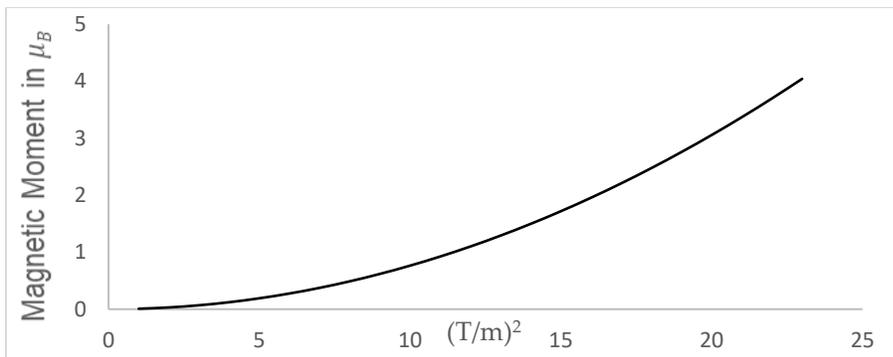

**Figure 3b: Magnitude of Muon Magnetic Moment (in units of $\mu_B$) vs $\left(\frac{T}{m}\right)^2$**

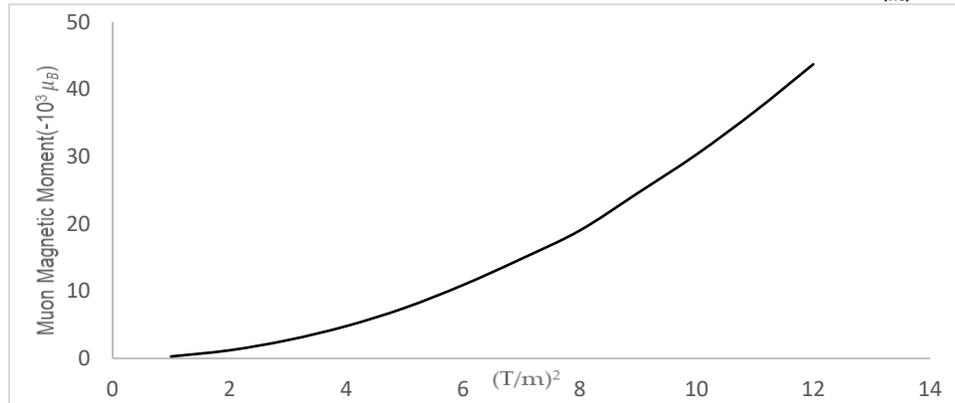

**Figure 3c: Magnitude of Tauon Magnetic Moment vs $\left(\frac{T}{m}\right)^2$**

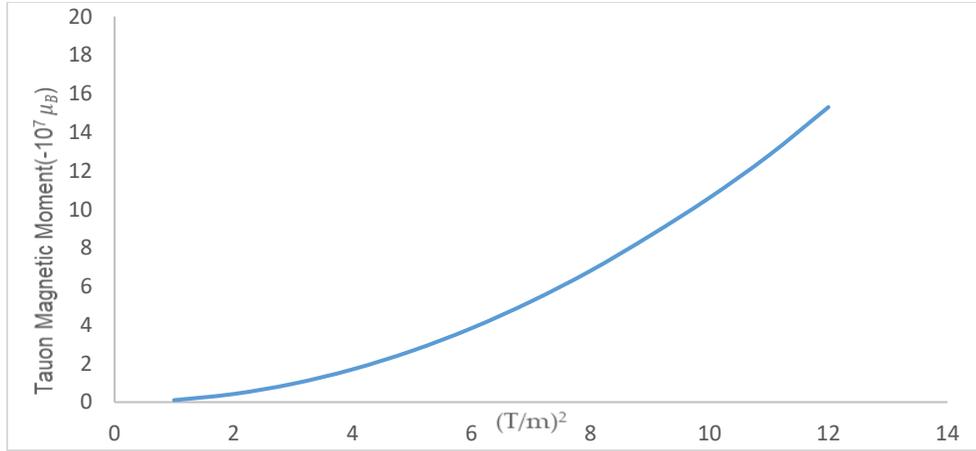

Fig. (4) shows how the magnetic moment flips when the temperature gets close to 0.22. This flip will occur for muons and tauons at much higher temperatures. Thermal contributions to the magnetic moment for the muon and tauon particles are too small to make a difference near nucleosynthesis and the concentration of heavier leptons in the universe was ignorable. In Figs. (3b) and (3c), we see similar magnetic moment behavior of muons and tauons as that of electrons, but their contribution will not be ignorable at larger T. However, electron magnetic moment calculations will need to incorporate weak interactions at larger temperatures. This figure gives a comparison of thermal contributions to the magnetic moment of all three charged leptons. Solid line corresponds to the magnetic moment of electron whereas the dotted line correspond to the muon magnetic moment. The dashed line indicates thermal contribution to the magnetic moment of tauons.

**Figure 4: Comparison of Magnetic Moment of various flavors vs $\left(\frac{T}{m}\right)^2$**

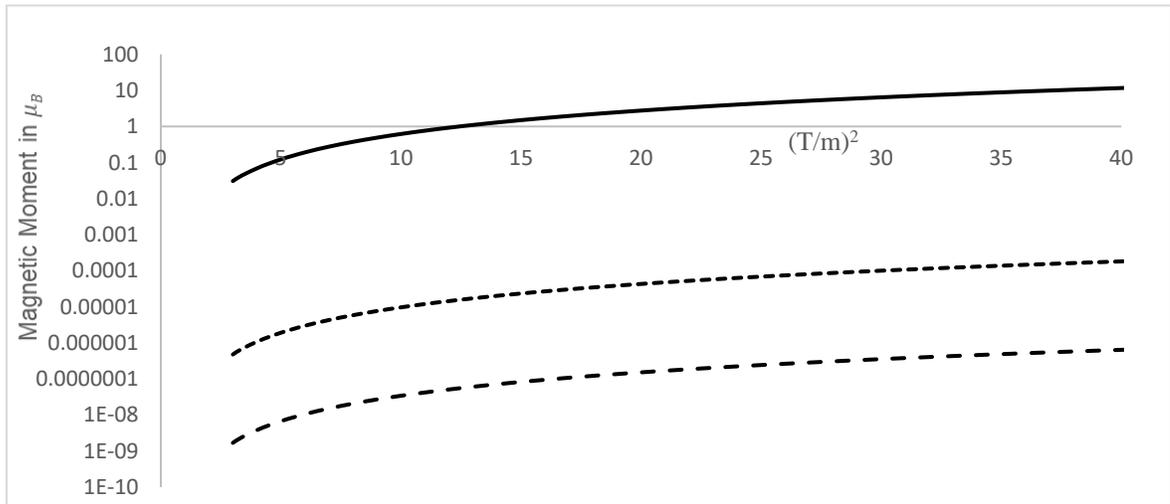

Leading order thermal contribution to leptons near the nucleosynthesis temperature (T ~ $m_e$) is compared in Table 1 to clearly show the relevance of different contributions at certain temperatures. This is not the exact value as we did not include abc functions contributions. The abc functions can contribute slightly to electron magnetic moment but the order of magnitude comparison is unchanged. Moreover, these functions are more relevant to show regular change in magnetic moment due to the change in temperature and composition during nucleosynthesis.

Neutrino masses and the associated magnetic moment are all very small but they are relevant to understand the particle behavior in the early universe. It is known that the magnetic moment of neutrino does not help to resolve the solar neutrino problem. However, its importance cannot be denied in the early universe and inside stars. This is especially true in superdense stars like supernova and neutron stars where the exact calculations of properties of particles are measurable physical parameters.

Table 2 gives a comparison of contributions of various flavors of neutrino magnetic moment around nucleosynthesis temperature. It gives a quantitative analysis of magnetic moment values near the nucleosynthesis temperatures corresponding to the given values of neutrino masses. These results can easily be modified if the mass limits are further improved in latest experiments. Current experimental limit from KATRIN [25] is showing the lower mass of neutrino so we just include the mass of $\nu_e$ at 1eV for comparison.

**Table 2: Magnetic Moment of Neutrinos**

| Neutrino Flavor | Mass $(eV)$ | Corresponding Temperature $(K)$ | Magnetic Moment at T=0 $\mu_B$ | Magnetic Moment with Thermal Contribution at T=$m_e$ $\mu_B$ |
|---|---|---|---|---|
| $\nu_e$ | 1 | $1.16 \times 10^4$ | $3.2 \times 10^{-19}$ | $2.92 \times 10^{-16}$ |
| $\nu_e$ | 2.25 | $2.6 \times 10^4$ | $7.2 \times 10^{-19}$ | $6.58 \times 10^{-16}$ |
| $\nu_\mu$ | $1.9 \times 10^5$ | $2.2 \times 10^9$ | $6.08 \times 10^{-14}$ | $5.55 \times 10^{-11}$ |
| $\nu_\tau$ | $1.55 \times 10^7$ | $2.1 \times 10^{11}$ | $4.96 \times 10^{-12}$ | $4.53 \times 10^{-9}$ |

We have plotted the magnetic moment of all flavors of neutrinos to compare their contributions in Fig. (5), for the same range of temperatures around T ~ m. The solid line, with minimum contribution, represents the electron type neutrino, the dotted line gives a plot of muon type neutrino, and the dashed line corresponds to tauon neutrino mass. Thermal contributions to the magnetic moment corresponding to all three generation of lepton doublets give a relevance of contribution of neutrino mass for different flavors. It also shows that the variation with temperature is exactly similar in each case, just the value of magnetic moment is inversely proportional to mass.

**Figure 5: Neutrino Magnetic Moment vs $\left(\frac{T}{m}\right)^2$**

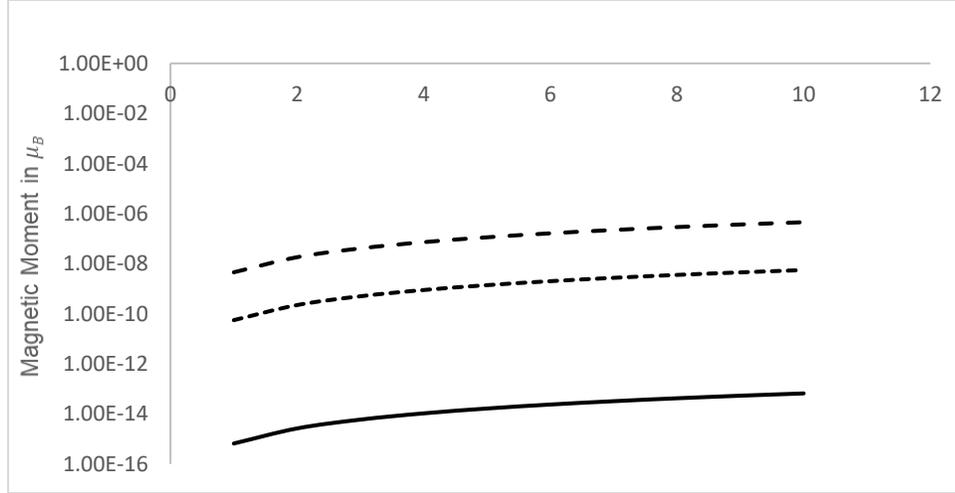

The graphs in Figs. (6) show the relationship between T/m and magnetic moment of individual neutrino flavor. In the beginning it appears to be linear, but its quadratic behavior becomes significant at higher temperatures. These graphs gives a comparison of thermal correction in reference to their masses. A comparison of thermal contribution to neutrino magnetic moment for various flavors of neutrino depends quadratically on temperature expressed in units of mass of charged lepton of the same generation. It is also worth-mentioning that the magnetic moment contribution is mainly related to the bubble diagram (Fig.1). On the other hand, contributions of tadpole diagrams (Fig.2), based on neutral current is fully suppressed.

**Figure 6a: Electron Neutrino Magnetic Moment vs $\left(\frac{T}{m}\right)^2$**

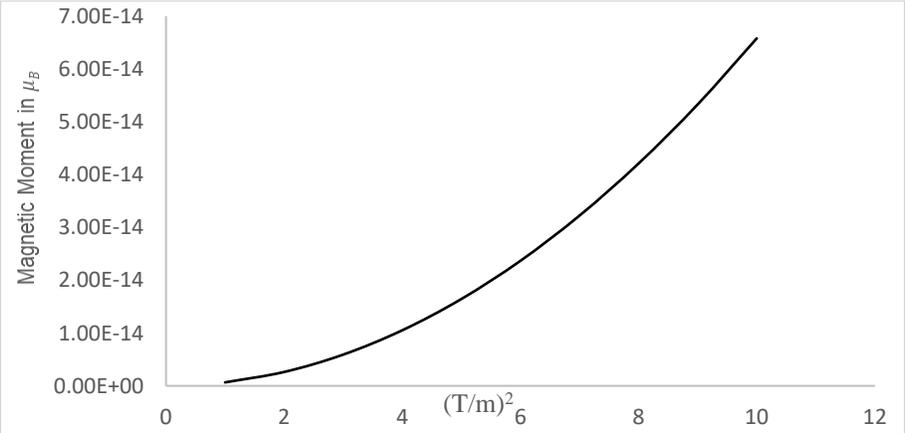

**Figure 6b: Muon Neutrino Magnetic Moment vs $\left(\frac{T}{m}\right)^2$**

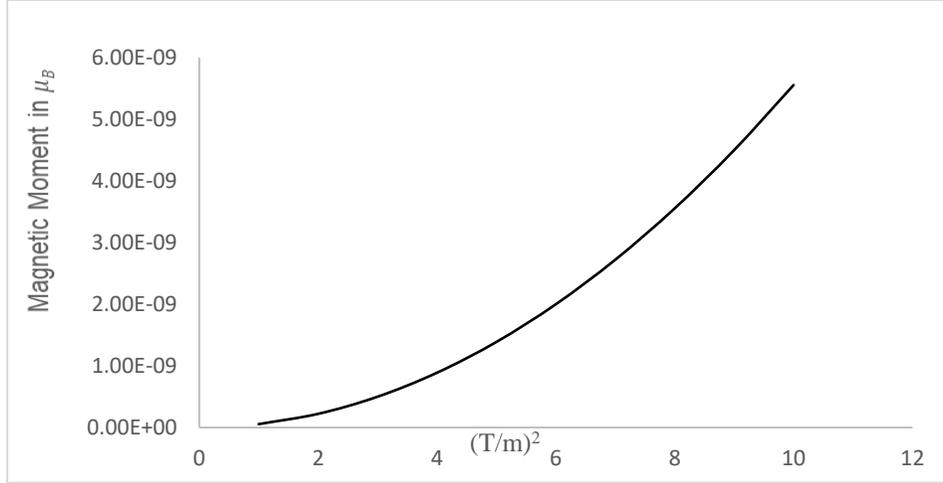

Figure 6c: Tau Neutrino Magnetic Moment vs $\left(\frac{T}{m}\right)^2$

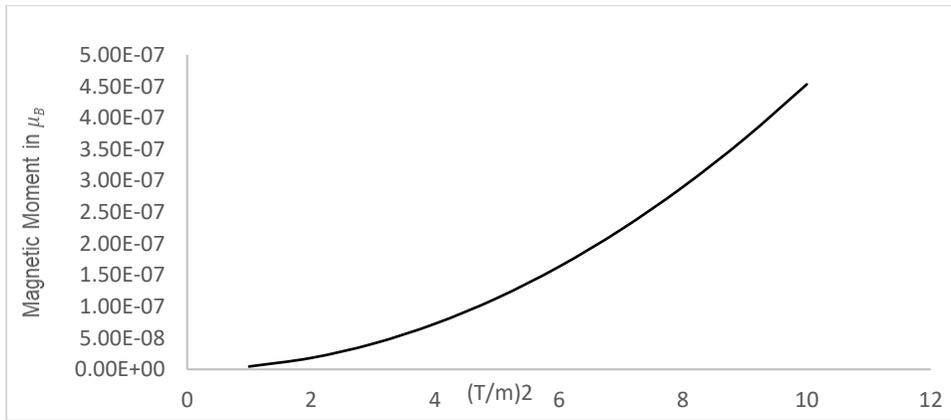

We have considered one set of masses of neutrinos, just as an example. However, these masses are subject to corrections through improved techniques of experimental measurements and theoretical models but the ratio among masses of various flavors is almost similar in every model. However, neutrino interactions with their media are model dependent and discussion of different models is out of scope of this paper.

The quantitative analysis of chemical potential contributions to the magnetic moments are not included in the discussion because the chemical composition and chemical potential changes are not well-known in any stellar model for superdense systems. There is also a lot of missing information about the phase of matter in stellar systems. The general expressions of Eq. (13) helps to understand the contribution of charged lepton. Chemical potential of neutrinos is not important due to its ignorable mass compared to the charged lepton mass of the same generation.

5. **Summary and Conclusion**

Magnetic moment of leptons is calculated using the straightforward generalization of the electron magnetic moment and the corresponding neutrino. We compare the statistical contribution of magnetic moment of various lepton flavors at different temperatures and chemical potentials and show how the relevance of this contribution depends on the statistical conditions of the background. For this purpose, the temperature and chemical potential are expressed in units of the corresponding lepton mass. Lower temperature and chemical potential contribute more to light leptons and the corresponding neutrino magnetic moment more than the heavier ones as shown in Figures (3) and (4). Since this contribution is induced by the selfmass of leptons given in Eqs.(11) and depends on various statistical conditions.

Renormalization scheme of QED is used to compute these corrections. A comparative study in Table 1 shows that the thermal correction on electron mass are fully estimated by QED only below the decoupling temperature [5] because of the low mass of electron. Whereas, for heavier leptons, QED corrections are still relevant for muons for temperature greater than the muon mass that is close to 200 times the electron mass. Tauon mass is very large and the relevance to the corresponding temperature can be considered around those temperatures or chemical potentials. However thermal contribution is much smaller for heavier leptons as compared to lighter ones.

Neutrino magnetic moment is calculated using the electroweak coupling of neutrinos. The background contribution of neutrino magnetic moment is related to the QED type coupling of charge with the external magnetic field. However, the masses of neutrinos are not measured. So we have given a comparison for a model dependent mass to show how the magnetic moment of different species of neutrinos is compared. However the mass of neutrinos is sufficiently smaller than the corresponding charged lepton mass of the same flavor showing the smaller magnetic moment contribution. A comparison of different neutrino flavors is given in Table 2.

Acknowledgements: The authors will like to thank Dr. Serkan Caliskan for reading the paper and some helpful suggestions.